\documentclass[11pt]{jpp}
\usepackage{epstopdf, epsfig}

\newcommand{\be}{\begin{displaymath}}
\newcommand{\bn}{\begin{equation}}
\newcommand{\bea}{\begin{eqnarray*}}
\newcommand{\eea}{\end{eqnarray*}}
\newcommand{\en}{\end{equation}}
\newcommand{\ee}{\end{displaymath}}

\renewcommand{\p}{\partial}
\newcommand{\lang}{\left\langle}
\newcommand{\rang}{\right\rangle}

\shorttitle{Available energy}
\shortauthor{P. Helander}

\title{Available energy and ground states of collisionless plasmas}

\author{Per Helander\aff{1}
  \corresp{\email{per.helander@ipp.mpg.de}}
}

\affiliation{\aff{1}Max-Planck-Institut f\"ur Plasmaphysik, D-17491 Greifswald, Germany
}

\begin{document}

\maketitle

\begin{abstract}
The energy budget of a collisionless plasma subject to electrostatic fluctuations is considered, and the 
excess of thermal energy over the minimum accessible to it under various constraints that limit the possible forms of plasma motion is calculated. This excess measures how much thermal energy is ``available'' for conversion into plasma instabilities, and therefore constitutes a nonlinear measure of plasma stability. A distribution function with zero available energy defines a ``ground state'' in the sense that its energy cannot decrease by any linear or nonlinear plasma motion. 
In a Vlasov plasma with small density and temperature fluctuations, the available energy is proportional to the mean square of these quantities, and exceeds the corresponding energy in ideal or resistive magnetohydrodynamics. If the first or second adiabatic invariant is conserved, ground states generally have inhomogeneous density and temperature. Magnetically confined plasmas are usually not in any ground state, but certain types of stellarator plasmas are so with respect to fluctuations that conserve both these adiabatic invariants, making the plasma linearly and nonlinearly stable to such fluctuations. Similar stability properties can also be enjoyed by plasmas confined by a dipole magnetic field. 
\end{abstract}

\section{Introduction}

Magnetised plasmas are subject to a plethora of instabilities. Some of these are catastrophic in the sense that they lead to a sudden loss of confinement or plasma termination, as in tokamak disruptions [\cite{Schuller-1995}] or solar flares [\cite{Fletcher-2011}], but most instabilities are of a more benign nature. In particular those with short wavelengths (microinstabilities) usually saturate at a low level and merely cause turbulent fluctuations and degradation of the confinement due to transport [\cite{Garbet-2004}]. 

The traditional way of mathematically investigating plasma stability is to identify an equilibrium state and then calculate the various linear eigenmodes and their growth rates. In simple cases, this can be done analytically, but accurate mode structures and growth rates can often only be found numerically. Great efforts have gone into the construction of computer codes for this purpose, see e.g.~\cite{Kotschenreuther-1995}. 

In reality, however, linear microinstabilities are rarely observed in laboratory and space plasmas. These are instead maintained in a nonlinearly fluctuating state, and the relevance of linear stability theory often seems questionable. The amplitude of the observed fluctuations do indeed seem to be related to the ``drives'' identified by linear stability theory, such as density and temperature gradients or unfavourable magnetic curvature [\cite{Garbet-2010}], but the assumptions made in linear stability theory are so restrictive that they rarely apply in practice. A question that arises, then, is whether there is some other way of characterising how ``unstable'' a plasma is, without resorting to any linear approximation. 

It is the purpose of the present paper to suggest a quantity that naturally lends itself to this purpose, which we shall call the ``available energy'' of the plasma. It is related to the concept of ``available potential energy'' in meteorology, which was introduced by \cite{Lorenz-1955} and denotes the excess of potential energy of the atmosphere above the minimum that could result from any adiabatic redistribution of mass. Analogously, we shall consider the excess of thermal energy of a magnetically confined plasma over the minimum accessible to it under the various constraints that limit the possible forms of plasma motion. This excess measures how much thermal energy is ``available'' for conversion into linear and nonlinear plasma instabilities, and therefore constitutes an intuitive measure of plasma stability. Of particular interest are plasmas with zero available energy. Such a plasma is in its lowest possible energy state, the ``ground state'', and is therefore stable to linear and nonlinear perturbations. 

The available energy is in general much smaller than the total thermal energy, and depends on what type of instability is considered -- or rather on what constraints are imposed on the motion of the plasma. For instance, if we consider collisionless, electrostatic instabilities with frequencies below the ion cyclotron frequency in a plasma embedded in a constant magnetic field of strength $B$, the magnetic moment $\mu=mv_\perp^2/2B$ is conserved for all plasma particles, and the average perpendicular pressure
	$$ \lang p_\perp \rang = \lang \int \mu B f \; d^3v \rang $$
therefore cannot change. (Here $f$ is the distribution function and angular brackets denote a volume average.) The part of the thermal energy that is associated with particle motion perpendicular to the magnetic field is therefore not available to drive such instabilities. This condition was recently used to derive upper bounds on magnetic-field generation by dynamo action in collisionless plasmas [\cite{Helander-2016-c}].

Another important constraint was identified by \cite{Gardner-1963} and applies if the distribution function is governed by the Vlasov equation, 
	\bn \frac{\p ({\sqrt{g}} \; f)}{\p t} 
	+ \nabla \cdot \left( \sqrt{g} \; \dot{\bf x} f \right) = 0, 
	\label{kinetic eq}
	\en
where ${\bf x}$ denotes arbitrary phase-space coordinates, $\nabla = {\p}/{\p {\bf x}}$, and $\sqrt{g}$ is the Jacobian, which satisfies Liouville's theorem,
	\bn 
	\frac{\p ({\sqrt{g}})}{\p t} + \nabla \cdot \left( \sqrt{g} \; \dot{\bf x} \right) = 0. 
	\label{Liouville}
	\en
Multiplying the Vlasov equation by $G'(f)$ and integrating over $\bf x$ gives
	\bn \frac{d}{dt} \lang \int G(f) \; d^3v \rang = 0 
	\label{integral constraint}
	\en
for any function $G$. There is thus an uncountable infinity of constraints on the evolution of $f$, and Gardner went on to determine the distribution function $f_0(v)$ that minimises the energy under these constraints. The purpose of the present work is to extend his calculation by accounting for additional constraints and apply it to magnetically confined plasmas. 

In the following section, we define the available energy for any kinetic plasma model that satisfies a Liouville theorem. In Section 3 we calculate the available energy of a Vlasov plasma and show that it is larger than the corresponding quantity in ideal and resistive MHD in Section 4. In the next two sections, we consider the effect of conservation laws such as magnetic-moment conservation, and find that the minimum-energy state in general is inhomogeneous unless the magnetic field is constant over the plasma volume. Finally, in Section 7, we explore the effect of the second adiabatic invariant and show that its conservation implies that plasmas confined in certain magnetic configurations have particularly small available energy. 

\section{The ground state}

Let us consider the evolution of the distribution function of some plasma species, assuming that it satisfies the Vlasov equation or any kinetic equation of the form (\ref{kinetic eq}) with the property that Liouville's theorem (\ref{Liouville}) holds. The question we are asking is: what is the minimum energy
	\bn E = \int \epsilon f \sqrt{g} \;  d{\bf x} 
	\label{E}
	\en
accessible to the distribution function? Here $\epsilon({\bf x})$ denotes the energy (usually $mv^2/2$) of an individual particle at position $\bf x$ in phase space. Clearly, the lowest possible value for $E$ would be attained if the support of $f$ (i.e., the region in which it assumes non-zero values) were concentrated to the point(s) where $\epsilon({\bf x})$ vanishes. The energy $E$ would then also vanish, but this state is usually not accessible from the initial condition. 

If we denote the energy-minimising distribution function (the ``ground state'') by $f_0({\bf x})$, then it should not be possible for the system to evolve away from this state in such a way that $E$ decreases. Now, if $f({\bf x},t_0) = f_0({\bf x})$, then for small $\delta t$ we have 
	$$ f({\bf x}, t_0+\delta t) - f({\bf x}, t_0) \simeq \delta t \frac{\p f}{\p t}
	= \delta {\bf y} \cdot \nabla f_0, $$
where $\delta {\bf y} = - \dot{\bf x}({\bf x},t_0) \delta t$. The variation in energy is thus
	$$ \delta E = E(t_0+\delta t) - E(t_0) = 
	\int \epsilon (\delta {\bf y} \cdot \nabla f_0 ) \sqrt{g} \;  d{\bf x}
	+ {\cal O}(\delta t^2). $$
and should vanish to first order in $\delta t$. This will clearly be the case if the functional $\delta E[\delta {\bf y}]$ vanishes for {\em all} trial functions $\delta {\bf y}$ such that $\nabla \cdot (\delta {\bf y} \sqrt{g}) = 0$. We thus introduce a Lagrange multiplier $\lambda({\bf x})$ and 
demand from $f_0$ that it should make the functional
	\bn \delta W = \int \left[ \epsilon (\delta {\bf y} \cdot \nabla f_0 ) \sqrt{g} 
	+ \lambda \nabla \cdot (\delta {\bf y} \sqrt{g}) \right] \;  d{\bf x}. 
	\label{dW}
	\en
vanish for all functions $\delta \bf y$ that vanish at infinity (or other boundaries of phase space). This is the case if, and only if, 
	$$ \epsilon \nabla f_0 = \nabla \lambda, $$
which implies that $f_0$ must be a function of $\epsilon$ alone, $f_0({\bf x}) = F_0[\epsilon({\bf x})]$. This can, for instance, be seen by differentiating the last equation, giving
	$$ (\p_i \epsilon) (\p_j f_0) - (\p_j \epsilon) (\p_i f_0) = 0 $$
for all index pairs $(i,j)$, where we have written $\p_i = \p / \p x_i$. 

To determine the function $F_0(\epsilon)$, we appeal to Eq.~(\ref{integral constraint}) with $G(f) = \Theta(f - \phi)$, where $\Theta$ denotes the Heaviside step function and $\phi$ an arbitrary constant, and we thus conclude that the function
	\bn H(\phi) = \int \Theta[f({\bf x}) - \phi] \sqrt{g} \; d{\bf x} 
	\label{H}
	\en
must equal 
	\bn \int \Theta[F_0(\epsilon({\bf x})) - \phi] \sqrt{g} \; d{\bf x}.  
	\label{H(F0)}
	\en
This statement is just a reformulation of Liouville's theorem: since the motion in phase space is incompressible, the volume in which the distribution exceeds a certain value, $\phi$, cannot change with time, whatever the choice of $\phi$. Furthermore, since $F_0(\epsilon)$ is a decreasing function of energy (see below), Eq.~(\ref{H(F0)}) must be equal to 
	\bn \int \Theta\left[\epsilon_\phi - \epsilon({\bf x}) \right] \sqrt{g} \; d{\bf x}
	= \Omega(\epsilon_\phi), 
	\label{Omega}
	\en
where $\epsilon_\phi$ denotes the energy for which $F_0(\epsilon_\phi) = \phi$. In other words, $\epsilon_\phi$ is the inverse of the function $F_0(\epsilon)$ and
	$$ \Omega'(y) = \int \delta[\epsilon({\bf x}) - y] \sqrt{g} \; d{\bf x} 
	$$
is the ``density of states'' of energy $y$. We thus conclude that the ground state is determined by the integral equation
	\bn H\left[ F_0(\epsilon) \right] = \Omega(\epsilon),
	\label{ground state}
	\en
which is fundamental to what follows. Any monotonically decreasing function of energy, $f({\bf x}) = F[\epsilon({\bf x})]$ with $dF/d\epsilon < 0$, is a possible ground state, because such a function trivially satisfies Eq.~(\ref{ground state}). If this equation is differentiated, we obtain an 
integro-differential equation
	$$ \frac{dF_0}{d\epsilon} = \frac{\Omega'(\epsilon)}{H'[F_0(\epsilon)]}, 
	$$
with the boundary condition $F_0(\epsilon) \rightarrow 0$ as $\epsilon \rightarrow \infty$. This equation has earlier been derived by \cite{Dodin-2005} for the slightly more special case that the trajectories are determined by Hamiltonian equations of motion. We prefer to work with Eq.~(\ref{ground state}) since it represents the integral of the Dodin-Fisch equation.\footnote{Note that this equation implies that $d F_0 / d \epsilon \le 0$, as assumed, since $H' \le 0 \le \Omega'$.}

Once either of these equations has been solved, the energy of the ground state can be calculated from
	$$ E_0 = \int \epsilon F_0 \sqrt{g} \;  d{\bf x}, $$
and the available energy, $A$, defined to be the difference between the energy (\ref{E}) of the initial state $f$ and that of the ground state,
	\bn A = E - E_0. 
	\label{A}
	\en
This is the maximum energy available for conversion into nonlinear fluctuations in the plasma. Strictly speaking, it is however only an upper bound on this energy, because we have not accounted for constraints other than that implied by Liouville's theorem. For instance, we have not considered whether the final state, with kinetic energy in fluctuations, is actually accessible from the initial condition. \cite{Fisch-1993} and \cite{Hay-2015} explored the maximum extractable energy under diffusion by waves, and found that it is generally less than that given by Eq.~(\ref{A}).

\section{Available energy of Vlasov plasmas}

Having defined the concepts of ground states and available energy, we now consider a couple of particularly simple examples where the evolution of the plasma is governed by the Vlasov equation with no additional constraints. The ground state will then always have constant density and temperature, since $f_0$ is function of energy alone. Conversely, if the initial  plasma density or temperature, defined by
	$$ n({\bf r}) = \int f({\bf r}, {\bf v}) d{\bf v}, $$
	\bn T({\bf r}) = \frac{2}{3n({\bf r})} \int \epsilon({\bf v}) f({\bf r}, {\bf v}) d{\bf v}, 
	\label{T}
	\en
in Cartesian phase-space coordinates ${\bf x} = ({\bf r}, {\bf v})$, varies over the plasma volume, then the plasma is not in a ground state and the available energy will in general be non-zero. 

Since $\epsilon = m v^2/2$, the integrated density of states (\ref{Omega}) is
	\bn \Omega(y) = \int d{\bf r} 
	\int_0^\infty \Theta \left( y  - \frac{m v^2}{2}\right) 4 \pi v^2 dv
	= \frac{4 \pi V}{3} \left(\frac{2y}{m}\right)^{3/2}, 
	\label{Cartesian Omega}
	\en
where $V$ is the volume of the spatial domain. This function plays a fundamental role in statistical mechanics and has recently stood in the centre of a debate on the correct definition of entropy [\cite{Dunkel-2014}].

\subsection{Bi-Maxwellian initial condition} 

Our first example is a bi-Maxwellian distribution function, for simplicity taken to be constant in space, 
	\bn f({\bf v}) = M e^{- \frac{m v_\perp^2}{2 T_\perp} - \frac{m v_\|^2}{2 T_\|}}, 
	\label{bi-Maxwellian}
	\en
where $M= n (m/2\pi \bar T)^{3/2}$ with $\bar T=T_\perp^{2/3} T_\|^{1/3}$. The energy density associated with this distribution function is
	$$ \frac{E}{V} = \frac{3nT}{2}, $$
where $T = (2 T_\perp + T_\|)/3$ denotes the temperature defined in Eq.~(\ref{T}). 

To calculate the ground state, we note that the surface $f({\bf v})=\phi$ in velocity space is given by 
	$$ \frac{m v_\perp^2}{2 T_\perp} + \frac{m v_\|^2}{2 T_\|} = \ln \left( \frac{M}{\phi} \right), $$
and represents an ellipsoid that encloses the volume
		$$ H(\phi) = \frac{4 \pi V}{3} \left( \frac{2 \bar T}{m} \ln \frac{M}{\phi} \right)^{3/2}. $$
The ground-state equation (\ref{ground state}) thus becomes
		$$ \frac{4 \pi V}{3} \left( \frac{2 \bar T}{m} \ln \frac{M}{F_0} \right)^{3/2}
		= \frac{4 \pi V}{3} \left(\frac{2\epsilon}{m}\right)^{3/2}, $$
where we have used Eq.~(\ref{Cartesian Omega}), and the solution is simply a Maxwellian,
	\bn F_0(\epsilon) = M e^{-\epsilon / \bar T}. 
	\label{ground state of a bi-Maxwellian}
	\en
The available energy is thus 
	\bn A = \frac{3 n V}{2} \left( T - \bar T \right), 
	\label{bi-Maxwellian A}
	\en
and is always positive, since $T$ denotes the arithmetic mean between the temperatures $(T_\perp,T_\perp,T_\|)$ in the three directions of velocity space, which always exceeds the geometric mean $\bar T$. 

This example clearly illustrates the difference between equilibrium states and ground states of the Vlasov equation. A spatially constant bi-Maxwellian is an equilibrium solution of this equation, but the ground state is Maxwellian, even in the absence of collisions.

\cite{Fowler-1968} used thermodynamic arguments, specifically the Helmholtz free energy, to derive bounds on the fluctuation amplitude in turbulent plasmas. \cite{Schekochihin-2017} has shown that when his technique is applied to a bi-Maxwellian, the result is the same as Eq.~(\ref{bi-Maxwellian A}).

\subsection{Maxwellian initial condition} 

Our second example is an initially Maxwellian distribution function,
	\bn f_M({\bf r}, {\bf v}) = n({\bf r}) \left( \frac{m}{2 \pi T({\bf r})} \right)^{3/2}
	e^{-m v^2 / 2 T({\bf r})}, 
	\label{fM}
	\en
whose density and temperature are now allowed to vary with $\bf r$. The total thermal energy is
	\bn E_M = \frac{3 V}{2} \lang n T \rang, 
	\label{EM}
	\en
where angular brackets again denote a spatial average. If we denote the maximum (over $\bf v$) of the Maxwellian (\ref{fM}) by 
	\bn M({\bf r}) = f_M({\bf r}, 0) 
	= n({\bf r}) \left(\frac{m}{2 \pi T({\bf r})} \right)^{3/2}, 
	\label{M}
	\en
the function (\ref{H}) becomes
	$$ H(\phi) = \frac{4 \pi}{3} \int \left( \frac{2 T}{m} \ln \frac{M}{\phi} \right)^{3/2} 
	\; \Theta(M-\phi) \; d{\bf r}, $$
and the ground-state equation (\ref{ground state})
	\bn \epsilon^{3/2} = 
	\lang \left[ T({\bf r}) \ln \frac{M({\bf r})}{F_0(\epsilon)} \right]^{3/2}
	\; \Theta[M({\bf r})-F_0(\epsilon)] \rang.
	\label{ground state solution}
	\en
Note that the ground state $F_0(\epsilon)$ determined by this nonlinear integral equation is generally not Maxwellian. 

We are now in a position to calculate the ground-state energy
	$$ \frac{E_0}{V} = \int_0^\infty \epsilon F_0 4 \pi v^2 dv 
	= \frac{4 \pi}{5} \left(\frac{2}{m} \right)^{3/2} \int_0^{\max F_0} \epsilon^{5/2} dF_0, $$
where we substitute the solution (\ref{ground state solution}) and obtain
	\bn \frac{E_0}{V} = \frac{4 \pi}{5} \left(\frac{2}{m} \right)^{3/2}
	\int_0^{\max F_0} \lang \left(T \ln \frac{M}{F_0} \right)^{3/2} \Theta( M - F_0) \rang^{5/3} dF_0. 
	\label{E0}
	\en
According to Jensen's inequality,	we have
	$$ \lang \left(T \ln \frac{M}{F_0} \right)^{3/2} \Theta( M - F_0) \rang^{5/3}
	\le \lang \left(T \ln \frac{M}{F_0} \right)^{5/2} \Theta( M - F_0) \rang, $$
and it follows that 
	$$ E_0 \le E_M, $$
with equality if, and only if, $n({\bf r})$ and $T({\bf r})$ are constant over the domain in question. As expected, a Maxellian (\ref{fM}) with spatially varying density or temperature thus has higher energy (\ref{EM}) than the ground state. 	
	
A general formula for the difference -- the available energy -- can be derived if the density and temperature only vary slightly, so that we can write 
	$$ n({\bf r}) = \lang n \rang [1 + \nu({\bf r}) ], $$
	$$ T({\bf r}) = \lang T \rang [1 + \tau({\bf r}) ], $$
with $\nu \sim \tau \ll 1$. Then the energy of the Maxwellian (\ref{EM}) becomes
	\bn E_M = \frac{3V}{2} \lang n \rang \lang T \rang \lang 1 + \nu \tau \rang, 
	\label{EM1}
	\en
and the function (\ref{M})
	$$ M = \bar{M} (1+\nu) (1 + \tau)^{-3/2}, $$
where
	$$ \bar M	= \lang n \rang \left(\frac{m}{2 \pi \lang T \rang} \right)^{3/2}. $$
The ground-state energy (\ref{E0}) can now be expressed as
	\bn \frac{E_0}{V} = \frac{4}{5\sqrt{\pi}} \lang n \rang \lang T \rang
	\int_{x_0}^\infty \lang (1+\tau)^{3/2}
	\left[x + \ln \frac{1+\nu}{(1+\tau)^{3/2}} \right]^{3/2} 
	\Theta \rang^{5/3} e^{-x} dx ,
	\label{Vlasov ground-state energy}
	\en
where $x = \ln (\bar M/F_0)$ and $x_0 = \ln (\bar M / \max F_0)$. Most of the contribution to the integral on the right-hand side comes from values of $x$ much larger than $\nu \sim \tau \ll 1$. The integral can therefore be calculated by expanding the integrand to second order in $\nu$ and $\tau$, and then integrating over $x$ from $0$ to $\infty$. Since
	$$ \ln \frac{1+\nu}{(1+\tau)^{3/2}} \simeq \nu - \frac{\nu^2}{2} - \frac{3 \tau}{2} + \frac{3 \tau^2}{4}, $$
and $\lang \nu \rang = \lang \tau \rang = 0$, we have
	$$ \lang (1+\tau)^{3/2}
	\left[x + \ln \frac{1+\nu}{(1+\tau)^{3/2}}  \right]^{3/2} \rang^{5/3} \simeq $$
	$$ x^{5/2} + \frac{5 x^{3/2}}{4} \lang 3 \nu \tau - {\nu^2} - 3 \tau^2 \rang
	+ \frac{5 x^{1/2}}{8} \lang \nu^2 - 3 \nu \tau + \frac{9 \tau^2}{4} \rang. 
	$$
which can be inserted into the integral (\ref{Vlasov ground-state energy}), giving
	$$ \frac{E_0}{V} = \frac{3V}{2} \lang n \rang \lang T \rang 
	\lang 1 - \frac{\nu^2}{3} - \frac{\tau^2}{2} + \nu \tau \rang.
	$$
Comparing with Eq.~(\ref{EM1}), we can thus express the available energy as
	\bn \frac{A}{E_M} = \lang \frac{\nu^2}{3} + \frac{\tau^2}{2} \rang. 
	\label{Vlasov A}
	\en
 
\subsection{Several particle species}

A simple consequence of Eq.~(\ref{Vlasov A}) is that it implies a limit on the amount of energy that can be transferred between different kinds of particles in a multi-species plasma. A common situation in astrophysics is that one particle species, usually the ions, has a significantly higher temperature than the other(s). A legitimate question, then, is whether there is any mechanism, linear or nonlinear, by which energy can be transferred from the hotter to the colder species.

Equation (\ref{Vlasov A}) implies an essentially negative answer to this question. In the absence of collisions, no energy can be extracted from a Maxwellian (or any other decreasing function of energy) unless there are spatial density and temperature variations. Only the available energy associated with the latter is accessible for extraction and transfer to other species. The temperature difference between different species does not itself contribute to the available energy. On the other hand, density and temperautre gradients may drive instabilities that indirectly lead to energy exchange between different species -- even if their temperatures happen to be equal [\cite{Barnes-2017}]. 
	
\section{Comparison with MHD}
	
It is interesting to compare the result (\ref{Vlasov A}) with the corresponding one obtained in magnetohydrodynamics (MHD) by \cite{Helander-2016-c}. From the continuity equation and the entropy conservation law of ideal MHD,
	$$ \frac{d \rho}{dt} + \rho \nabla \cdot {\bf V} = 0, $$
	$$ \frac{d}{dt} \left( \frac{p}{\rho^\gamma} \right) = 0, $$
where $\rho$ denotes the density, $p = nT$ the pressure, $\gamma = 5/3$ the adiabatic index, and $\bf V$ the fluid velocity, 
it is straightforward to derive the equation
	$$ \frac{\p p^{1/\gamma}}{\p t} + \nabla \cdot \left( p^{1/\gamma} {\bf V} \right) = 0. $$
If there is no flow across the boundary, it follows that 
	$$ S = \lang p^{1/\gamma} \rang $$ 
is a conserved quantity, which is a manifestation of the entropy law in the present context. If other constraints (e.g. topological ones) are ignored, the lowest-energy state accessible to an MHD plasma is obtained by minimsing the thermal energy 
	$$ E = \frac{\lang p \rang}{\gamma - 1} V, $$
at fixed $S$. The resulting state is one of constant pressure $p_0 = \lang p^{1/\gamma} \rang^\gamma$, and the available energy is thus given by [\cite{Helander-2016-c}]
	$$ A_{\rm MHD} = \frac{\lang p \rang- \lang p^{1/\gamma} \rang^\gamma}{\gamma-1}V. $$
If the initial pressure perturbations are small, $p({\bf r}) = \lang p \rang  + \delta p({\bf r})$, the available energy becomes
	$$ A_{\rm MHD} = \frac{\lang p \rang}{2 \gamma} 
	\lang \left( \frac{\delta p}{p} \right)^2 \rang V$$ 
to second order in the fluctuations. By making the identification $p=nT = \lang n \rang \lang T \rang ( 1 + \nu + \tau + \nu \tau )$, we can compare this result with Eq.~(\ref{Vlasov A}). The difference in available energies becomes
  $$ A - A_{\rm MHD} 
	= \frac{\lang p \rang}{5} \lang \left(\nu - \frac{3\tau}{2} \right)^2 \rang V \ge 0, $$
showing that more energy is available in a Maxwellian Vlasov plasma than in MHD. This conclusion does not change if resistivity and viscosity are introduced into the latter. The entropy then increases, causing $S$ to grow and $A_{\rm MHD}$ to drop. Note that $A = A_{\rm MHD}$ exactly when the relative fluctuations in density are everywhere 50\% larger than those in temperature, $\nu = 3 \tau/2$, implying that the linearised entropy perturbation vanishes,
	$$ \delta \ln \left(\frac{p}{\rho^\gamma} \right) = \frac{\delta p}{p} - \frac{\gamma \delta \rho}{\rho} = \tau - \frac{2\nu}{3} = 0. $$

\section{Conserved quantities}

The derivation given in Section 2 can easily be extended to cases where the motion is constrained to satisfy certain conservation laws. The ground states then become less trivial and the concept of available energy correspondingly more interesting. If some quantity such as the magnetic moment $\mu({\bf x})$ is conserved, then the trial function $\delta \bf y$ in Eq.~(\ref{dW}) must be required to satisfy $\delta {\bf y} \cdot \nabla \mu = 0$, which can be accounted for by using $\mu$ as one of the phase-space coordinates. We thus write ${\bf x} = ({\bf z}, \mu)$ and demand that the integral
	$$ \delta W[ \delta {\bf y} ; \mu ] 
	= \int \left[ \epsilon (\delta {\bf y} \cdot \nabla f_0 ) \sqrt{g} 
	+ \lambda \nabla \cdot (\delta {\bf y} \sqrt{g}) \right] \;  d{\bf z}, $$
should vanish for every value of $\mu$, which is to be held fixed in the integration. This condition implies that $f_0({\bf x})$ depends on the phase-space coordinates only through $\epsilon$ and $\mu$, i.e.~it must be a function of the form $f_0({\bf z}, \mu) = F_0[\epsilon({\bf z}, \mu), \mu]$. In analogy with the constraint (\ref{integral constraint}) we now have
	$$ \frac{d}{dt} \int G(f) \sqrt{g} \; d{\bf z} = 0 $$
for all $\mu$, which is again to kept constant in the integral. As in Eqs.~(\ref{H})-(\ref{Omega}), we define
	$$ H(\phi, \mu) = \int \Theta[f({\bf x}) - \phi] \sqrt{g} \; d{\bf z}, $$
	$$ \Omega(w,\mu) = 
	\int \Theta\left[w - \epsilon({\bf x}) \right] \sqrt{g} \; d{\bf z}, $$
and conclude that the ground state $F_0(\epsilon,\mu)$ is determined by the equation
	\bn H\left[ F_0(\epsilon,\mu), \mu \right] = \Omega(\epsilon, \mu). 
	\label{ground state with mu}
	\en

Of course, it is trivial to generalise these equations to the case of multiple conservation laws. One merely needs to replace the scalar $\mu$ by a vector representing all the conserved quantities. Note that any distribution function of the form $F_0(\epsilon,\mu)$ is a ground state if $\p F_0 / \p \epsilon \le 0$ everywhere. 
	
\section{Conservation of the magnetic moment}

We now consider the case where $\mu$ specifically denotes the magnetic moment, $\mu = m v_\perp^2/(2B)$, and demonstrate that the density and temperature of the ground state will in general not be constant but depend on the strength of the magnetic field $B({\bf r})$. 

A particularly simple example is furnished by a two-dimensional plasma with a magnetic field ${\bf B} = B(x,y)\nabla z$ pointing in the ignorable direction and varying in the two other directions. Any function $F_0(\mu)$ of $\mu$ alone is then a ground state, with density
	$$ n = \int_0^\infty F_0 \; 2 \pi v_\perp dv_\perp 
	= \frac{2 \pi B}{m} \int_0^\infty F_0 \; d\mu, $$
proportional to $B(x,y)$. In other words, there is no available energy associated with density variations proportional to $B$ if the magnetic moment is conserved. As a consequence, if the plasma is initially in a state of constant density but varying magnetic field strength, turbulence may be expected to spontaneously cause the density to become non-uniform [\cite{Yankov-1997}]. The temperature is
	$$ T = \frac{1}{n} \int_0^\infty \frac{mv_\perp^2}{2}
	F_0 \; 2 \pi v_\perp dv_\perp 
	= \frac{2 \pi B^2}{mn} \int_0^\infty \mu F_0 \; d\mu, $$
and will then also tend to a state proportional to $B(x,y)$. 

In three dimensions, any function $F_0(\epsilon, \mu)$ of energy $\epsilon = mv^2/2$ and magnetic moment is a ground state if $\p F_0/\p \epsilon \le 0$ [\cite{Taylor-1963}]. A simple example is the spatially homogeneous bi-Maxwellian (\ref{bi-Maxwellian}), which is a ground state if the magnetic field strength is constant but, as we have seen, is not a ground state if the magnetic moment is allowed to vary. The conservation of $\mu$ thus prevents access to the lower-energy state (\ref{ground state of a bi-Maxwellian}) otherwise available to the plasma. 

More generally, the density and temperature of a ground state $F_0(\epsilon, \mu)$,
 	$$ { n \choose T} = \pi B  \left( \frac{2}{m} \right)^{3/2} 
	\int_0^\infty  {1 \choose \frac{2\epsilon}{3n}} \; d\epsilon
	\int_0^{\epsilon/B} \frac{F_0(\epsilon, \mu) \, d \mu}{\sqrt{\epsilon - \mu B}}, $$
will be non-uniform (unless $\p F_0 / \p \mu = 0$) if the magnetic field is inhomogeneous. 

\section{The adiabatic invariant of parallel motion}

We now turn our attention to plasmas where the frequencies of collisions and any fluctuations are lower than that of particle motion along the magnetic field, so that not only the magnetic moment, but also the parallel adiabatic invariant,
	$$ J = \int mv_\| \, dl, $$
is conserved. Here $l$ denotes the arc length along $\bf B$, and the integration is carried out between two consecutive turning points of the motion. ($J$ is usually conserved for electrons in hot fusion plasmas, since the frequency of the turbulence is comparable to the diamagnetic frequency, which is smaller than that of the electron transit frequency.) Any distribution function of the form $F_0(\epsilon, \mu, J)$ is now a ground state if 
	$$ \left( \frac{\p F_0}{\p \epsilon} \right)_{\mu, J} \le 0 $$
for every choice of $\mu$ and $J$. This condition coincides with the linear stability criterion derived by \cite{Taylor-1964} for electrostatic flute modes in mirror machines in the zero-gyroradius-limit, but is of much wider significance [\cite{Schmidt-1965}, \cite{Taylor-2015}]. If this criterion is satisfied for all particle species, the plasma is not only linearly stable, but is in a global lowest-energy state and is therefore also nonlinearly stable. A detailed argument using gyrokinetics is given in the Appendix. 

In a tokamak or stellarator, it is useful to write the magnetic field as ${\bf B} = \nabla \psi \times \nabla \alpha$, where $\psi$ labels the toroidally nested flux surfaces and $\alpha$ the different field lines on each such surface.\footnote{Strictly speaking, the magnetic surfaces need not be nested here, but could also form islands. In general, ground states and available energy can be defined for {\em any} plasma state, whether in equilibrium or not, regardless of the nature of the magnetic field, which could even be chaotic.} The distribution function of each species is usually of the form of a Maxwellian whose density and temperature are constant on flux surfaces,
	\bn f_M(\psi, \epsilon) = n(\psi) \left[ \frac{m}{2 \pi T(\psi)} \right]^{3/2} e^{- \epsilon / T(\psi)}. 
	\label{tokamak fM}
	\en
Expressing $\psi$ as a function $\psi = \psi(\mu, J, \epsilon)$, we find
	$$ \left( \frac{\p f_M}{\p \epsilon} \right)_{\mu, J} 
	= \left( \frac{\p f_M}{\p \psi} \right)_{\epsilon} \left( \frac{\p \psi}{\p \epsilon} \right)_{\mu, J}
	- \frac{f_M}{T}, $$
where [\cite{Helander-2014-a}]
		\bn \left( \frac{\p \psi}{\p \epsilon} \right)_{\mu, J} 
		= - \left( \frac{\p J}{\p \epsilon} \right)_{\mu, \psi}
		\bigg\slash \left( \frac{\p J}{\p \psi} \right)_{\mu, \epsilon}
		= \frac{1}{q \omega_\alpha},  
		\label{precession}
		\en
$q$ denotes the electric charge, and $\omega_\alpha = \overline{{\bf v}_d \cdot \nabla \alpha}$ the frequency of trapped-particle precession in the $\alpha$-direction. Here ${\bf v}_d$ is the drift velocity and an overbar indicates an orbit average. Thus, in standard gyrokinetic notation,
	\bn \left( \frac{\p f_M}{\p \epsilon} \right)_{\mu, J} 
	= \frac{f_M}{T} \left( \frac{\omega_\ast^T}{\overline \omega_d}  - 1 \right),
	\label{dfM/de}
	\en
where $\omega_d = {\bf k}_\perp \cdot {\bf v}_d$ with ${\bf k}_\perp = k_\alpha \nabla \alpha$, 
	$$ \omega_\ast = \frac{k_{\alpha} T}{q} \frac{d \ln n}{d \psi}, $$
  $$ \omega_\ast^T = \omega_\ast \left[1 + \eta \left( \frac{\epsilon}{T} - \frac{3}{2} \right) \right], $$
and $\eta = (d \ln T / d \psi) / (d \ln n / d \psi)$. (The wave number $k_\alpha$ drops out and is only introduced for reasons of convention.) According to Eq.~(\ref{dfM/de}), the Maxwellian (\ref{tokamak fM}) is a ground state if 
	\bn \frac{\omega_\ast^T}{\overline \omega_d} < 1
	\label{stability condition}
	\en
for all values of $\epsilon$, $\mu$ and $J$. As we shall see in the next two subsections, there are two different ways to satisfy this condition. 

\subsection{Maximum-$J$ devices}

The criterion (\ref{stability condition}) holds if the following two conditions are both satisfied:
	\bn \omega_\ast \overline \omega_d \le 0, \label{max-J}
	\en
	\bn 0 \le \eta < \frac{2}{3}. \label{low eta}
	\en
The first of these conditions is the maximum-$J$ criterion, $\p J / \p \psi < 0$ if $d n / d \psi < 0$, which was identified by \cite{Rosenbluth-1968} as stabilising to low-frequency interchange modes. It has recently been applied to gyrokinetic instabilities in quasi-isodynamic stellarators by \cite{Proll-2012} and \cite{Helander-2013}, who showed that the two criteria (\ref{max-J}) and (\ref{low eta}) imply linear stability of collisionless trapped-particle modes. The physical significance of Eq.~(\ref{max-J}) is that it guarantees that all trapped particle particles experience favourable magnetic curvature on a time-average along their orbits. 

This is never the case in tokamaks. The magnetic curvature is unfavourable on the outboard side of the torus, where deeply trapped particles reside, whereas more shallowly trapped particles spend most of their time on the inboard side of the torus, where they experience good curvature. The drift frequency $\overline \omega_d$ is thus positive for some orbits and negative for others, so the distribution function (\ref{tokamak fM}) cannot possibly correspond to a ground state. 

The situation is different in certain stellarators, where the regions of particle trapping and bad magnetic curvature are separated from each other. This tends to be the case, at least to some approximation, in quasi-isodynamic stellarators at high beta. Thanks to the diamagnetic property of the plasma, its pressure digs a magnetic well surrounding the magnetic axis, causing the poloidal precession caused by the grad-$B$ drift to reverse relative to the diamagnetic frequency, so that $\omega_\ast \overline \omega_d$ becomes negative for all trapped orbits. If $0 \le \eta < 2/3$, the plasma is then in a ground state relative to fluctuations that conserve $\mu$ and $J$. This tends to be the case for the plasma electrons, since the typical turbulence frequency is of order $\omega_\ast$, which is much smaller than the electron bounce frequency if with $k_\perp$ is comparable to the inverse ion gyroradius [\cite{Isichenko-1996}]. For the ions, however, the bounce frequency does not exceed the turbulence frequency, and $J$ is not expected to be conserved. 

An interesting question that arises, then, is the nature of the lowest-energy state in a plasma where $J$ is conserved for the electrons but not for the ions. Different constraints then apply to the different particle species, which are however coupled to each other by the requirement of quasineutrality. The ground state of such a plasma should be sought by minimising the energy of both species simultaneously, each subject to its own particular constraints and additionally to that of quasineutrality. This problem is beyond the scope of the present publication. 

\subsection{Unfavourable magnetic curvature}

Remarkably, there is another way of satisfying the stability criterion (\ref{stability condition}) if the magnetic curvature is {\em unfavourable} for all orbits. The bounce-averaged drift frequency is proportional to the kinetic energy,
	$$ \overline \omega_d = \tilde \omega_d(\lambda,\psi,\alpha) \frac{\epsilon}{T(\psi)}, $$ 
where $\lambda = \mu / \epsilon$, and if $\tilde \omega_d \omega_\ast$ is positive for all trapped orbits, so that all such trajectories on average experience bad magnetic curvature, then Eq.~(\ref{stability condition}) is satisfied if simultaneously 
	\bn \eta > \frac{2}{3} 
	\label{large eta}
	\en
and 
	\bn \frac{\eta \omega_\ast}{\tilde \omega_d} < 1 
	\label{2nd condition}
	\en
for all particles. It is impossible to satisfy the second of these criteria in a tokamak, because $\tilde \omega_d$ has different signs for deeply and shallowly trapped orbits. It therefore goes through zero, making $\eta \omega_\ast / \tilde \omega_d$ arbitrarily large. 

In a dipole magnetic field, however, the magnetic curvature is bad everywhere, and the precession frequency is an increasing function of $\lambda$ [\cite{Kesner-2002}], bounded from below by
	$$ \tilde \omega_d(\lambda) \ge \tilde \omega_d(\lambda =0). $$
To calculate this bound, we note that for $\lambda = 0$,
		$$ J = mvL(\psi), $$
where $L(\psi)$ denotes the length of the field lines on the flux surface labelled by $\psi$. For a point dipole, ${\bf B} = \nabla \psi \times \nabla \alpha$, where $\alpha$ denotes the toroidal angle, we have
	$$ \psi = \frac{M}{r} \sin^2 \theta $$
in spherical coordinates. It follows that $L(\psi)$ is inversely proportional to $\psi$ and thus
	$$ \frac{\p J}{\p \psi} = - \frac{J}{\psi}, $$
	$$ \frac{\p J}{\p \epsilon} = \frac{J}{2 \epsilon}, $$
so that the drift frequency from Eq.~(\ref{precession}) becomes
	$$ \omega_d(\lambda =0) = \frac{2 k_\alpha \epsilon}{q \psi}. $$
According to Eqs.~(\ref{large eta}) and (\ref{2nd condition}), the plasma is thus in a ground state if
	\bn \frac{2}{3} \frac{d \ln n}{d \ln \psi} < \frac{d \ln T}{d \ln \psi} < 2,
	\label{dipole stability}
	\en
and nonlinear stability is then guaranteed. This region in parameter space occupies a subset of the linear stability domain identified in earlier works [\cite{Kesner-2002,Helander-2014,Helander-2016-b}], see Fig.~1. 

\begin{figure}
  \centerline{\includegraphics[width=0.7\textwidth]{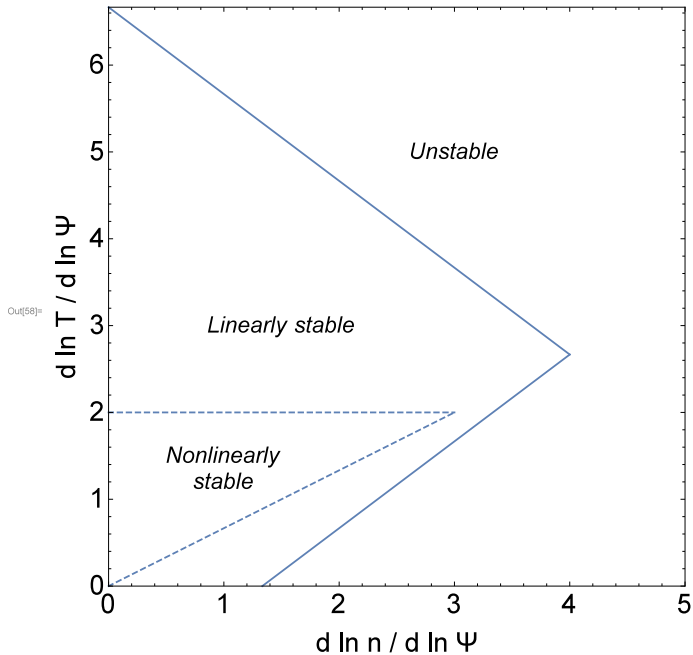}}
  \caption{Stability diagram for low-frequency electrostatic perturbations of a collisionless plasma in the magnetic field of a point dipole. The linear stability region was calculated under the simplifying assumption that the precession frequency is independent of the trapping parameter $\lambda$, which is not quite true but was shown by \cite{Kesner-2002} to be a good approximation. The nonlinear stability domain is calculated without this approximation. }
\label{fig1}
\end{figure}

\subsection{Partly unfavourable curvature}

It is, of course, impossible to realise a pure dipole magnetic field, and we therefore briefly turn our attention to the field produced by a circular coil of finite radius. In this case, the two ways of satisfying the ground-state criterion (\ref{stability condition}) we have just discussed can be combined in such a way that Eqs.~(\ref{max-J})-(\ref{low eta}) hold in some parts of the plasma and Eqs.~(\ref{large eta})-(\ref{2nd condition}) elsewhere. If $I$ denotes the current and $a$ the radius of the coil, the magnetic flux function becomes [\cite{Jackson-1975}]
	$$ \psi(r,\theta) = \frac{\mu_0 I a r \sin \theta}{\pi \sqrt{a^2 + r^2 + 2 ar \sin \theta}}
	\frac{(2-k^2) K(k) - 2 E(k)}{k^2}, $$
where $K$ and $E$ denote complete elliptic integrals of the argument
	$$ k^2 = \frac{4 a r \sin \theta}{a^2 + r^2 + 2 ar \sin \theta}. $$
The magnetic drift is everywhere in the same  toroidal direction, so $\omega_d$ never changes sign. The plasma density and temperature, however, should go to zero both close to the coil, where $\psi$ becomes large, and far away from it, where $\psi \rightarrow 0$. The density and temperature will therefore peak on some intermediate surface, at $\psi = \psi_0$ say, where the diamagnetic frequency thus changes sign. The product $\tilde \omega_d \omega_\ast^T$ is negative inside this surface ($\psi > \psi_0$), reflecting favourable magnetic curvature, whereas $\tilde \omega_d \omega_\ast^T > 0 $ and the curvature is unfavourable in the outer region ($\psi < \psi_0$). The ground-state condition (\ref{stability condition}) can therefore be satisfied throughout the entire plasma by choosing the density and temperature profiles so that $0 < \eta < 2/3$ in the inner region and $\eta > 2/3$ in the outer region. In the latter region, the temperature profile should also satisfy the condition $\eta \omega_\ast < \min \tilde \omega_d$, which far from the coil ($r \gg a$) reduces to Eq.~(\ref{dipole stability}). It thus appears possible to use a single (levitated) coil to confine a collisionless plasma in such a way that it is nonlinearly stable to all low-frequency, electrostatic perturbations. 

Work is currently underway to construct a such a device for confining electron-positron plasmas [T.S.~Pedersen, private communication (2017)]. The density will be so low that the Debye length exceeds the gyroradius by a large factor, and for this reason modes with frequencies above the bounce frequency are predicted to be linearly stable [\cite{Helander-2014}]. Given this linear stability at high frequencies and the nonlinear stability at lower ones, there is every reason to hope for good confinement. 

Excellent plasma performance in dipole magnetic fields has indeed been observed experimentally [\cite{Yoshida-2010,Boxer-2010}] and predicted theoretically on the basis of linear stability calculations [\cite{Simakov-2002}] or relaxation arguments [\cite{Hasegawa-1990}]. The latter, however, imply non-Maxwellian distribution functions that are unlikely to apply if the confinement time exceeds the collision time. 

\begin{figure}
  \centerline{\includegraphics[width=0.7\textwidth]{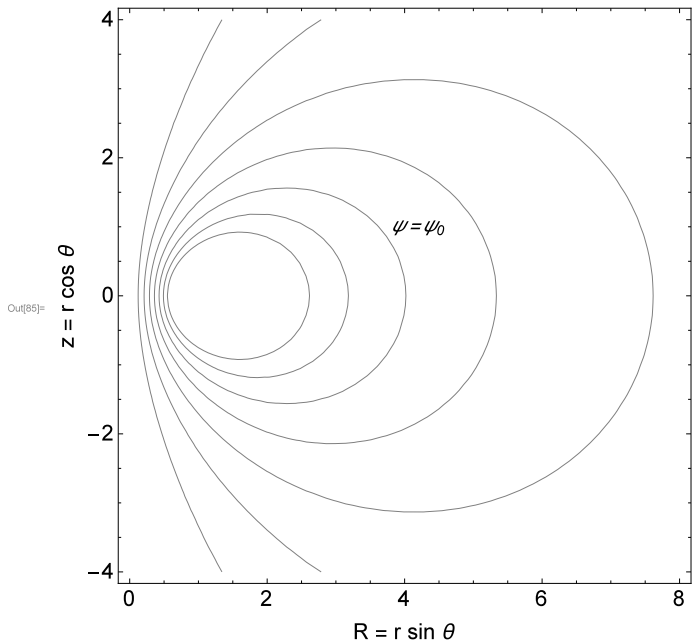}}
  \caption{Field lines produced by a circular coil at $R = 1, z = 0$. If the density and temperature peak on the field line labelled by $\psi = \psi_0$, the magnetic curvature is favourable inside this line and unfavourable outside it. The plasma is in a ground state if $0 < \eta < 2/3$ in the inner region and the gradients in the outer region lie in the nonlinearly stable region corresponding to Fig.~1.}
\end{figure}

\section{Conclusions}

We have defined a ground state of a plasma species as a state with the lowest possible energy subject to whatever constraints limit the possible motion. The available energy is the difference between the actual energy of the plasma and the ground-state energy. It provides an upper bound on how much thermal energy can be converted into instabilities or turbulence. 

If no constraint other than the Liouville theorem is imposed, the ground state of a Vlasov plasma is homogeneous, and the available energy of an initially Maxwellian plasma is proportional to the mean square of the density and temperature fluctuations, if these are assumed to be small. This energy exceeds the corresponding quantity calculated from ideal or visco-resistive MHD. The ground state of a spatially constant bi-Maxwellian distribution function is Maxwellian, even in the absence of collisions. 

If the first or second adiabatic invariant ($\mu$ or $J$) is conserved, the plasma parameters of most ground states vary over the plasma volume. If instabilities or turbulence bring the plasma toward such a state, spontaneous density and temperature peaking may thus result. 

In maximum-$J$ devices such as quasi-isodynamic stellarators, the average magnetic curvature is favourable for all particle trajectories and the plasma is in a ground state with respect to fluctuations that conserve $\mu$ and $J$ if the temperature profile is less than two thirds as steep as the density profile. The plasma is thus stable, even nonlinearly, to electrostatic instabilities with frequencies smaller than the bounce frequency of all species. Collisionless trapped-ion modes fall into this category, but the more common trapped-electron modes have frequencies that are comparable to the ion bounce frequency, so that $J$-conservation is broken for the ions. Since it still holds for the electrons, these act stabilising and any instability must draw its energy from the ions. This result has earlier been found through linear stability analysis by \cite{Proll-2012}, but is here found to be true even for large perturbations. 

Perhaps more surprisingly, the plasma can be in a ground state even in magnetic fields with {\em unfavourable} curvature, if the temperature profile is {\em more} than two thirds as steep as the density profile. For instance, in the field of a magnetic dipole, nonlinear stability to low-frequency electrostatic modes should prevail if Eq.~(\ref{dipole stability}) is satisfied. As a result, in appears that a low-density plasma could be confined extremely well by the field from a single levitated coil.

We have limited our discussion to collisionless and electrostatic instabilities, but it would be interesting also to explore the concepts of available energy and ground states in more general settings. 
%

\section*{Acknowledgment}

I would like to thank Gabriel Plunk, Nat Fisch, Alex Schekochihin and Ian Abel for many insightful comments and interesting discussions. 

\section*{Appendix: nonlinear stability}

On several occasions in this paper, it has been stated that ground states are stable, not only linearly but also to finite-sized perturbations. The precise meaning of this statement is that if the plasma is prepared in such a way that all distribution functions initially are in ground states, then the energy of the electric-field fluctuations, suitably defined, cannot grow with time, independently of their initial amplitude. In order to prove this assertion, it is necessary to consider the full system of equations describing the electric field as well as all plasma particle species. We shall do so for the Vlasov-Poisson and gyrokinetic systems of equations, respectively. In both cases, it is sufficient to show that a quantity of the form
	$$ W = E + \Phi $$
is conserved, where $\Phi$ denotes the energy of fluctuations, which must be required to be positive definite, and $E$ denotes the sum of all particle energies, taking the form
	\bn E = \sum_a \int \epsilon f_a \sqrt{g} \;  d{\bf x} 
	\label{sum of E}
	\en
for some suitable function $\epsilon({\bf x})$. It then follows that
	$$ \Phi = W-E \le W-E_0 $$
if $E \ge E_0$. Thus, if the system is initiated in a state with total energy $W$ and the particles in a ground state, the fluctuation energy $\Phi$ cannot grow. 

\subsection*{The Vlasov-Poisson system}

The Vlasov-Poisson system of equations for an arbitrary number of species (distinguished by a subscript $a$) in an electric field ${\bf E} = - \nabla \phi$, is 
	$$ \frac{\p f_a}{\p t} + {\bf v} \cdot \nabla f_a - \frac{q_a}{m_a} \nabla \phi \cdot \frac{\p f_a}{\p \bf v} = 0, $$
	$$ \nabla^2 \phi = - 4 \pi \sum_a q_a \int f_a \, d{\bf v}. $$
If there is no flux to infinity or across other boundaries, this system of equations conserves the energy
	$$ W = E + \Phi, $$
where
	$$ E = \sum_a \lang \int \frac{m_a v^2}{2} f_a \, d{\bf v} \rang $$
denotes the energy (\ref{E}) of the particles, and 
	$$ \Phi = \lang \frac{| \nabla \phi |^2}{8 \pi} \rang $$
that of the electric field. Since both energies are positive definite, the stability criterion formulated above is satisfied.

\subsection*{Gyrokinetic system of equations}

We now demonstrate nonlinear stability of a plasma with respect to low-frequency fluctuations if the gyro-averaged distribution functions of all species satisfy
	$$ \left( \frac{\p F_0}{\p \epsilon} \right)_{\mu, J} \le 0 $$
everywhere in phase space. To this end, we note that the nonlinear, electrostatic system of gyrokinetic equations conserves the energy [\cite{Dubin-1983}]
	$$ W = E+ \Phi, $$
where
	$$ E = \sum_a \int f_a({\bf R}, \mu, v_\|) \kappa \sqrt{g} \; d{\bf x} , $$
	$$ \kappa = \mu B + \frac{m_a v_\|^2}{2}, $$
and 
	$$ \Phi = \int \left( \frac{| \nabla \phi |^2}{8 \pi} 
	+ \sum_a \frac{m_a n_a}{2} \left| \frac{{\bf B} \times \nabla \phi}{B^2} \right|^2 \right) d{\bf r}, $$
denotes the fluctuation energy in the long-wavelength limit, which is appropriate for low-frequency instabilities. Here $\bf R$ denotes the gyro-centre position and $\sqrt{g} \, d{\bf x} = (2 \pi B_\|^\ast/m_a) \, d{\bf R} dv_\| d \mu$ the phase-space volume element, which can also be written as
	$$ \sqrt{g} \; d{\bf x} = \frac{4 \pi}{m_a^2 | v_\| |} dH d\mu d \psi d\alpha dl, $$
where  
	$$ H = \kappa + e_a \phi({\bf R}) $$
is the total (kinetic + potential) guiding-centre energy. The parallel adiabatic invariant is
	$$ J(\mu, H, \psi, \alpha) = \int \sqrt{2 m_a (H - \mu B - e_a \phi)} \; dl, $$
and its derivative with respect to $H$
	$$ \frac{\p J}{\p H} = \int \frac{dl}{| v_\| |} = \tau_b, $$
the bounce time. Hence we can write
	$$ E = \sum_a \frac{4 \pi}{m_a^2} \int \frac{\kappa f_a dl}{|v_\| | \tau_b} d\mu dJ d\psi d\alpha. $$

For slow fluctuations (compared with the bounce frequency), the distribution function is nearly independent of $l$, and thus
	$$ W = E + \Phi, $$
where
	$$ E = \sum_a \frac{4 \pi}{m_a^2} \int \epsilon f_a d\mu dJ d\psi d\alpha, $$
and $\epsilon(\mu, J, \psi, \alpha)$ is defined as the bounce average of the kinetic gyro-centre energy, 
	$$ \epsilon = \frac{1}{\tau_b} \int \kappa \frac{dl}{|v_\||}. $$
With this definition of the particle energy $\epsilon({\bf x})$ in Eq.~(\ref{sum of E}), which is equal to the local kinetic energy if the electrostatic potential is a flux function (i.e., only depends on $\psi$), the conditions for nonlinear stability are thus satisfied. 

\newpage
\bibliographystyle{jpp}

\bibliography{jpp-per}

\begin{thebibliography}{34}
\expandafter\ifx\csname natexlab\endcsname\relax\def\natexlab#1{#1}\fi
\def\au#1{#1} \def\ed#1{#1} \def\yr#1{#1}\def\at#1{#1}\def\jt#1{\textit{#1}}
  \def\bt#1{#1}\def\bvol#1{\textbf{#1}} \def\vol#1{#1} \def\pg#1{#1}
  \def\publ#1{#1}\def\arxiv#1{#1}\def\org#1{#1}\def\st#1{\textit{#1}}

\bibitem[Barnes {\em et~al.\/}(2017)Barnes, Abiuso \& Dorland]{Barnes-2017}
{\sc \au{Barnes, M.}, \au{Abiuso, P.} \& \au{Dorland, W.}} \yr{2017}
  \at{Turbulent heating in an inhomogeneous, magnetized plasma slab}.
  \jt{submitted to J. Plasma Phys.} .

\bibitem[Boxer {\em et~al.\/}(2010)Boxer, Bergmann, Ellsworth, Garnier, Kesner,
  Mauel \& Woskov]{Boxer-2010}
{\sc \au{Boxer, A.~C.}, \au{Bergmann, R.}, \au{Ellsworth, J.~L.}, \au{Garnier,
  D.~T.}, \au{Kesner, J.}, \au{Mauel, M.~E.} \& \au{Woskov, P.}} \yr{2010}
  \at{Turbulent inward pinch of plasma confined by a levitated dipole magnet}.
  \jt{Nature Phys.}  \bvol{6},  \pg{207--212}.

\bibitem[Dodin \& Fisch(2005)]{Dodin-2005}
{\sc \au{Dodin, I.Y.} \& \au{Fisch, N.J.}} \yr{2005}  \at{Variational
  formulation of gardner's restacking algorithm}.  \jt{Phys. Lett. A}
  \bvol{341}~(1-4),  \pg{187--192}.

\bibitem[Dubin {\em et~al.\/}(1983)Dubin, Krommes, Oberman \& Lee]{Dubin-1983}
{\sc \au{Dubin, D.H.E.}, \au{Krommes, J.A.}, \au{Oberman, C.} \& \au{Lee,
  W.W.}} \yr{1983}  \at{Nonlinear gyrokinetic equations}.  \jt{Phys. Fluids}
  \bvol{26}~(12),  \pg{3524--3535}.

\bibitem[Dunkel \& Hilbert(2014)]{Dunkel-2014}
{\sc \au{Dunkel, J.} \& \au{Hilbert, S.}} \yr{2014}  \at{Consistent
  thermostatistics forbids negative absolute temperatures}.  \jt{Nature Phys.}
  \bvol{10}~(1),  \pg{67--72}.

\bibitem[Fisch \& Rax(1993)]{Fisch-1993}
{\sc \au{Fisch, N.J.} \& \au{Rax, J.-M.}} \yr{1993}  \at{Free energy in plasmas
  under wave-induced diffusion}.  \jt{Phys. Fluids B}  \bvol{5}~(6),
  \pg{1754--1759}.

\bibitem[Fletcher {\em et~al.\/}(2011)Fletcher, Dennis, Hudson, Krucker,
  Phillips, Veronig, Battaglia, Bone, Caspi, Chen, Gallagher, Grigis, Ji, Liu,
  Milligan \& Temmer]{Fletcher-2011}
{\sc \au{Fletcher, L.}, \au{Dennis, B.R.}, \au{Hudson, H.~S.}, \au{Krucker,
  S.}, \au{Phillips, K.}, \au{Veronig, A.}, \au{Battaglia, M.}, \au{Bone, L.},
  \au{Caspi, A.}, \au{Chen, Q.}, \au{Gallagher, P.}, \au{Grigis, P.~T.},
  \au{Ji, H.}, \au{Liu, W.}, \au{Milligan, R.~O.} \& \au{Temmer, M.}} \yr{2011}
   \at{An observational overview of solar flares}.  \jt{Space Science Reviews}
  \bvol{159},  \pg{19--106}.

\bibitem[Fowler(1968)]{Fowler-1968}
{\sc \au{Fowler, T.K.}} \yr{1968}  \at{Thermodynamics of unstable plasmas}.
  \jt{Advances in plasma physics. Volume 1. Edited by A. Simon and W. B.
  Thompson.}  \bvol{1},  \pg{201--225}.

\bibitem[Garbet {\em et~al.\/}(2010)Garbet, Idomura, Villard \&
  Watanabe]{Garbet-2010}
{\sc \au{Garbet, X.}, \au{Idomura, Y.}, \au{Villard, L.} \& \au{Watanabe,
  T.H.}} \yr{2010}  \at{Gyrokinetic simulations of turbulent transport}.
  \jt{Nucl. Fusion}  \bvol{50},  \pg{043002}.

\bibitem[Garbet {\em et~al.\/}(2004)Garbet, Mantica, Angioni, Asp, Baranov,
  Bourdelle, Budny, Crisanti, Cordey \& Garzotti]{Garbet-2004}
{\sc \au{Garbet, X.}, \au{Mantica, P.}, \au{Angioni, C.}, \au{Asp, E.},
  \au{Baranov, Y.}, \au{Bourdelle, C.}, \au{Budny, R.}, \au{Crisanti, F.},
  \au{Cordey, G.} \& \au{Garzotti, L.}} \yr{2004}  \at{Physics of transport in
  tokamaks}.  \jt{Plasma Phys. Control. Fusion}  \bvol{46},  \pg{B557}.

\bibitem[Gardner(1963)]{Gardner-1963}
{\sc \au{Gardner, C.S.}} \yr{1963}  \at{Bound on the energy available from a
  plasma}.  \jt{Phys. Fluids}  \bvol{6}~(6),  \pg{839--840}.

\bibitem[Hasegawa {\em et~al.\/}(1990)Hasegawa, Chen \& Mauel]{Hasegawa-1990}
{\sc \au{Hasegawa, A.}, \au{Chen, Liu} \& \au{Mauel, M.E.}} \yr{1990}  \at{A
  deuterium-helium-3 fusion reactor based on a dipole magnetic field}.
  \jt{Nucl. Fusion}  \bvol{30}~(11),  \pg{2405}.

\bibitem[Hay {\em et~al.\/}(2015)Hay, Schiff \& Fisch]{Hay-2015}
{\sc \au{Hay, M.J.}, \au{Schiff, J.} \& \au{Fisch, N.J.}} \yr{2015}
  \at{Maximal energy extraction under discrete diffusive exchange}.  \jt{Phys.
  Plasmas}  \bvol{22}~(10),  \pg{102108}.

\bibitem[Helander(2014{\natexlab{{\em a\/}}})]{Helander-2014}
{\sc \au{Helander, P.}} \yr{2014{\natexlab{{\em a\/}}}}  \at{Microstability of
  magnetically confined electron-positron plasmas}.  \jt{Phys. Rev. Lett.}
  \bvol{113},  \pg{135003}.

\bibitem[Helander(2014{\natexlab{{\em b\/}}})]{Helander-2014-a}
{\sc \au{Helander, P.}} \yr{2014{\natexlab{{\em b\/}}}}  \at{Theory of plasma
  confinement in non-axisymmetric magnetic fields}.  \jt{Rep. Prog. Phys.}
  \bvol{77}~(8),  \pg{087001}.

\bibitem[Helander \& Connor({2016})]{Helander-2016-b}
{\sc \au{Helander, P.} \& \au{Connor, J.W.}} \yr{{2016}}  \at{{Gyrokinetic
  stability theory of electron-positron plasmas}}.  \jt{{J. Plasma Phys.}}
  \bvol{{82}}~({3}).

\bibitem[Helander {\em et~al.\/}(2013)Helander, Proll \& Plunk]{Helander-2013}
{\sc \au{Helander, P.}, \au{Proll, J.H.E.} \& \au{Plunk, G.G.}} \yr{2013}
  \at{Collisionless microinstabilities in stellarators. i. analytical theory of
  trapped-particle modes}.  \jt{Phys. Plasmas}  \bvol{20}~({12}),  \pg{122505}.

\bibitem[Helander {\em et~al.\/}(2016)Helander, Strumik \&
  Schekochihin]{Helander-2016-c}
{\sc \au{Helander, P.}, \au{Strumik, M.} \& \au{Schekochihin, A.A.}} \yr{2016}
  \at{Constraints on dynamo action in plasmas}.  \jt{J. Plasma Phys.}
  \bvol{82},  \pg{905820601}.

\bibitem[Isichenko {\em et~al.\/}(1996)Isichenko, Gruzinov, Diamond \&
  Yushmanov]{Isichenko-1996}
{\sc \au{Isichenko, M.B.}, \au{Gruzinov, A.V.}, \au{Diamond, P.H.} \&
  \au{Yushmanov, P.N.}} \yr{1996}  \at{Anomalous pinch effect and energy
  exchange in tokamaks}.  \jt{Phys. Plasmas}  \bvol{3}~(5),  \pg{1916--1925},
  37th Annual Meeting of the Division-of-Plasma-Physics of the
  American-Physical-Society Location: LOUISVILLE, KY Date: NOV 06-10, 1995.

\bibitem[Jackson(1975)]{Jackson-1975}
{\sc \au{Jackson, J.D.}} \yr{1975}  \at{Classical electrodynamics, 2nd edition}
  .

\bibitem[Kesner \& Hastie(2002)]{Kesner-2002}
{\sc \au{Kesner, J.} \& \au{Hastie, R.J.}} \yr{2002}  \at{Electrostatic drift
  modes in a closed field line configuration}.  \jt{Phys. Plasmas}
  \bvol{9}~(2),  \pg{395--400}.

\bibitem[Kotschenreuther {\em et~al.\/}(1995)Kotschenreuther, Rewoldt \&
  Tang]{Kotschenreuther-1995}
{\sc \au{Kotschenreuther, M.}, \au{Rewoldt, G.} \& \au{Tang, W.M.}} \yr{1995}
  \at{Comparison of initial value and eigenvalue codes for kinetic toroidal
  plasma instabilities}.  \jt{Comp. Phys. Comm.}  \bvol{88}~(2-3),
  \pg{128--140}.

\bibitem[Lorenz(1955)]{Lorenz-1955}
{\sc \au{Lorenz, E.N.}} \yr{1955}  \at{Available potential energy and the
  maintenance of the general circulation}.  \jt{Tellus}  \bvol{7}~(2),
  \pg{157--167}.

\bibitem[Proll {\em et~al.\/}({2012})Proll, Helander, Connor \&
  Plunk]{Proll-2012}
{\sc \au{Proll, J.H.E.}, \au{Helander, P.}, \au{Connor, J.W.} \& \au{Plunk,
  G.G.}} \yr{{2012}}  \at{{Resilience of quasi-isodynamic stellarators against
  trapped-particle instabilities}}.  \jt{{Phys. Rev. Lett.}}
  \bvol{{108}}~({24}),  \pg{{245002}}.

\bibitem[Rosenbluth(1968)]{Rosenbluth-1968}
{\sc \au{Rosenbluth, M.N.}} \yr{1968}  \at{Low-frequency limit of interchange
  instability}.  \jt{Phys. Fluids}  \bvol{11},  \pg{869--872}.

\bibitem[Schekochihin(2017)]{Schekochihin-2017}
{\sc \au{Schekochihin, A.A.}} \yr{2017}  \at{Lecture notes on kinetic theory
  and magnetohydrodynamics of plasmas}.  \jt{Oxford University} .

\bibitem[Schmidt(1965)]{Schmidt-1965}
{\sc \au{Schmidt, G.}} \yr{1965}  \at{Extended stability criterion for
  minimum-b geometries}.  \jt{Phys. Fluids}  \bvol{8}~(4),  \pg{754--754}.

\bibitem[Schuller(1995)]{Schuller-1995}
{\sc \au{Schuller, F.C.}} \yr{1995}  \at{Disruptions in tokamaks}.  \jt{Plasma
  Phys. Control. Fusion}  \bvol{37},  \pg{A135}.

\bibitem[Simakov {\em et~al.\/}(2002)Simakov, Hastie \& Catto]{Simakov-2002}
{\sc \au{Simakov, A.N.}, \au{Hastie, R.J.} \& \au{Catto, P.J.}} \yr{2002}
  \at{Long mean-free path collisional stability of electromagnetic modes in
  axisymmetric closed magnetic field configurations}.  \jt{Phys. Plasmas}
  \bvol{9},  \pg{201}.

\bibitem[Taylor(1963)]{Taylor-1963}
{\sc \au{Taylor, J.B.}} \yr{1963}  \at{Some stable plasma equilibria in
  combined mirror-cusp fields}.  \jt{Phys. Fluids}  \bvol{6}~(11),
  \pg{1529--1536}.

\bibitem[Taylor(1964)]{Taylor-1964}
{\sc \au{Taylor, J.B.}} \yr{1964}  \at{Equilibrium and stability of plasma in
  arbitrary mirror fields}.  \jt{Phys. Fluids}  \bvol{7}~(6),  \pg{767--773}.

\bibitem[Taylor \& Newton(2015)]{Taylor-2015}
{\sc \au{Taylor, J.B.} \& \au{Newton, S.L.}} \yr{2015}  \at{Special topics in
  plasma confinement}.  \jt{J. Plasma Phys.}  \bvol{81},  \pg{205810501}.

\bibitem[Yankov \& Nycander(1997)]{Yankov-1997}
{\sc \au{Yankov, V.V.} \& \au{Nycander, J.}} \yr{1997}  \at{Description of
  turbulent transport in tokamaks by invariants}.  \jt{Phys. Plasmas}
  \bvol{4}~(8),  \pg{2907--2919}.

\bibitem[Yoshida {\em et~al.\/}(2010)Yoshida, Saitoh, Morikawa, Yano, Watanabe
  \& Ogawa]{Yoshida-2010}
{\sc \au{Yoshida, Z.}, \au{Saitoh, H.}, \au{Morikawa, J.}, \au{Yano, Y.},
  \au{Watanabe, S.} \& \au{Ogawa, Y.}} \yr{2010}  \at{Magnetospheric vortex
  formation: Self-organized confinement of charged particles}.  \jt{Phys. Rev.
  Lett.}  \bvol{104},  \pg{235004}.

\end{thebibliography}
	
\end{document}